%% file: main.tex
\documentclass[11pt]{article}
\usepackage{fullpage}
\usepackage{graphicx}
\usepackage{amsmath}

\usepackage{subcaption}
\usepackage{afterpage}

\usepackage{amssymb}
\usepackage[top=1in,bottom=1in,left=1in,right=1in]{geometry}

\usepackage{amsbsy}
\usepackage{mathtools}

\usepackage{natbib}

\usepackage{multirow}
\usepackage{booktabs}
\usepackage{amsbsy}
\usepackage{mathtools}
\usepackage{microtype}
\usepackage[table]{xcolor}
\usepackage{hhline}

\usepackage{dsfont}
\newcommand{\1}[1]{\mathds{1}\left[#1\right]}
\providecommand{\abs}[1]{\lvert#1\rvert}

\providecommand{\nth}[1]{{#1}^{\textrm{th}}}

\makeatletter
  \newcommand\scripty{\@setfontsize\scripty{6pt}{7}}
\makeatother

\usepackage{hyperref}
\hypersetup{colorlinks=true,linkcolor=black,citecolor=black,filecolor=black,urlcolor=black,pagebackref=true,breaklinks=true,bookmarks=true}

\begin{document}

\title{Alignment of dynamic networks}


\author{Vipin Vijayan $^{\text{1}}$, Dominic Critchlow
  $^{\text{1,2}}$, and Tijana Milenkovi\'{c} $^{\text{1,}*}$
}

\date{}
\maketitle

\begin{abstract}
\noindent{\bf Motivation: }Networks can model real-world systems in a
variety of domains. Network
alignment (NA) aims to find a node mapping that conserves similar
regions between compared networks. NA is applicable to many fields,
including computational biology, where NA can guide the transfer of
biological knowledge from well- to poorly-studied species across
aligned network regions. Existing NA methods can only align static
networks. However, most complex real-world systems evolve over time
and should thus be modeled as dynamic networks. We hypothesize that
aligning dynamic network representations of evolving systems will
produce superior alignments compared to aligning the systems' static
network representations, as is currently done.
\\{\bf Results: }For this purpose, we introduce the first ever dynamic NA method,
DynaMAGNA++. This proof-of-concept dynamic NA method is an extension
of a state-of-the-art static NA method, MAGNA++. Even though both
MAGNA++ and DynaMAGNA++ optimize edge as well as node conservation
across the aligned networks, MAGNA++ conserves static edges and
similarity between static node neighborhoods, while DynaMAGNA++
conserves dynamic edges (events) and similarity between evolving node
neighborhoods. For this purpose, we introduce the first ever measure
of dynamic edge conservation and rely on our recent measure of dynamic
node conservation. Importantly, the two dynamic conservation measures
can be optimized using any state-of-the-art NA method and not just
MAGNA++. We confirm our hypothesis that dynamic NA is superior to
static NA, under fair comparison conditions, on synthetic and
real-world networks, in computational biology and social network
domains. DynaMAGNA++ is parallelized and it includes a user-friendly
graphical interface.
\\
\textbf{Software:} Available upon request.\\
\textbf{Supplementary information:} Available upon request.\\
\textbf{Contact:} \href{tmilenko@nd.edu}{tmilenko@nd.edu},
\href{vvijayan@nd.edu}{vvijayan@nd.edu}
\end{abstract}
\makeatletter{\renewcommand*{\@makefnmark}{}\footnotetext{$^{\text{1}}$Department of Computer Science and
Engineering, ECK Institute for Global Health, and Interdisciplinary
Center for Network Science and Applications (iCeNSA), University of
Notre Dame, Notre Dame, IN 46556, USA\makeatother}
\makeatletter{\renewcommand*{\@makefnmark}{}\footnotetext{$^{\text{2}}$Department
  of Physics and Astronomy, Austin Peay State University, Clarksville, Tennessee, TN 37044, USA\makeatother}
\makeatletter{\renewcommand*{\@makefnmark}{}\footnotetext{$^\ast$To whom correspondence should be addressed.\makeatother}
\section{Introduction}
\label{sec:intro}

Networks can be used to model complex real-world systems in a variety
of domains \citep{ComplexNetworks}. Network alignment (NA)
compares networks with the goal of finding a node mapping that
conserves topologically or functionally similar regions between the
networks. NA has been used
in many domains and applications \citep{FiftyYears}. In computer
vision, it has been used to find correspondences between sets of
visual features \citep{NApaperDuchenne}.  In online social networks,
NA has been used to match identities of people who have different
account types (e.g., Twitter and Facebook) \citep{COSNET}.
In ontology matching, NA has been used to match concepts
across ontological networks \citep{NApaperBayatiJ}.  Computational biology is no
exception.  In this domain, NA has been used to predict protein
function (including the role of proteins in aging), by aligning
protein interaction networks (PINs) of different species, and by
transferring functional knowledge from a well-studied species to a
poorly-studied species between the species' conserved (aligned) PIN
regions
\citep{FaisalAging,ConeReview,ElmsallatiReview,LocalVsGlobal,GuzziNA}. Also,
NA has been used to construct phylogenetic trees of species based on
similarities of their PINs or metabolic networks
\citep{GRAAL,MIGRAAL}.

NA methods can be categorized as local or global
\citep{LocalVsGlobal,GuzziNA}.  Local NA typically finds
\emph{highly conserved} but consequently \emph{small} regions among
compared networks, and it results in a \emph{many-to-many} node
mapping. On the other hand, global NA typically finds a
\emph{one-to-one} node mapping between compared networks that results
in \emph{large} but consequently \emph{suboptimally conserved} network
regions. Clearly, each of local NA and global NA has its
(dis)advantages \citep{LocalVsGlobal,IGLOO,GuzziNA}.  In this paper,
we focus on global NA, but our ideas are applicable to local NA as
well. Also, NA methods can be categorized as pairwise or multiple
\citep{ConeReview,GuzziNA,multiMAGNA++}.  Pairwise NA aligns two networks while multiple NA
can align more than two networks at once.  While multiple NA can capture
conserved network regions between more networks than pairwise NA,
which may lead to deeper biological insights compared to pairwise NA,
multiple NA is computationally much harder than pairwise NA since the
complexity of the NA problem typically increases exponentially with
the number of networks. This is why in this paper we focus on pairwise
NA, but our ideas can be extended to multiple NA as well.  Henceforth,
we refer to global and pairwise NA simply as NA.

Existing NA methods can only align static networks.  This is because
in many domains and applications, static network representations are
often used to model complex real-world systems, independent of whether
the systems are static or dynamic. However, most real-world systems
are dynamic, as they evolve over time.  Static networks cannot fully
capture the temporal aspect of evolving systems. Instead, such systems
can be better modeled as dynamic networks \citep{ModernTemporal}.  For
example, a complex system such as a social network evolves over time
as friendships are made and lost. Static networks cannot model the
changes in interactions between nodes over time, while dynamic
networks can capture the times during which the friendships begin and
end.  Other examples of systems that can be more accurately
represented as dynamic networks include communication systems, human
or animal proximity interactions, ecological
systems, and many systems in biology that evolve over time,
including brain or cellular functioning.
In particular, regarding the latter, while cellular functioning is
dynamic, current computational methods (including all existing NA
methods) for analyzing \emph{systems-level} molecular networks, such
as PINs, deal with the networks' static representations. This is in
part due to unavailability of experimental dynamic molecular network
data, owing to limitations of biotechnologies for data
collection. Yet, as more dynamic molecular (and other real-world)
network data are becoming available, there is a growing need for
computational methods that are capable of analyzing dynamic networks
\citep{NetworkIntegrationDynamics,DynamicInteractome}, including methods that can align
such networks.

The question is: how to align dynamic networks, when the existing NA
methods can only deal with static networks? To allow for this, we
generalize the notion of static NA to its dynamic counterpart. Namely,
we define {\em dynamic NA} as a process of comparing dynamic networks and
finding similar regions between such networks, while exploiting the
temporal information explicitly (unlike static NA, which ignores this information). We hypothesize that
aligning dynamic network representations of evolving real-world systems
will produce superior alignments compared to aligning the systems'
static network representations, as is currently done. To test this
hypothesis, we introduce the first ever method for dynamic NA.

Our proposed dynamic NA method, DynaMAGNA++, is a proof-of-concept extension of a
state-of-the-art static NA method, MAGNA++ \citep{MAGNA++}.
\citet{MAGNA} and \citet{MAGNA++} compared MAGNA++ to state-of-the-art
static NA methods at the time, namely IsoRank \citep{IsoRank}, MI-GRAAL \citep{MIGRAAL},
and GHOST \citep{GHOST}. More recently, \citet{LocalVsGlobal} compared
MAGNA++ to additional newer static NA methods: NETAL \citep{NETAL},
GEDEVO \citep{GEDEVO}, WAVE \citep{WAVE}, and L-GRAAL \citep{LGRAAL}.
The comparisons were made on synthetic as well as real-world PINs, in terms of both
topological and functional alignment quality. MAGNA++ was found to be
superior to six of the seven existing methods and comparable to the remaining
method. This is exactly why we have chosen to extend MAGNA++ rather
than some other static NA method to its dynamic counterpart. However,
as any future static NA methods are developed \citep{SANA} that are potentially
superior to MAGNA++, our ideas on dynamic NA will be applicable to
such methods too.  Section \ref{sec:methods} describes the method, and
Section \ref{sec:results} confirms our hypothesis that dynamic NA is
superior to static NA, under fair comparison conditions, on both
synthetic and real-world networks, and on data from both computational
biology and social network domains.

\section{Methods}
\label{sec:methods}

We first summarize MAGNA++, and then we describe our proposed dynamic NA method,
DynaMAGNA++, as an extension of MAGNA++.

\subsection{MAGNA++}
\label{sec:magna}

\noindent {\bf Static networks and static NA.} A static network $G(V,E)$ consists of a
node set $V$ and an edge set $E$.  An edge $(u,v) \in E$ is an
interaction between nodes $u$ and $v$. There can only be a single edge
between the same pair of nodes.  Given two static networks
$G_1(V_1,E_1)$ and $G_2(V_2,E_2)$, assuming without loss of generality
that $|V_1| \le |V_2|$, a static NA between $G_1$ and $G_2$ is a
one-to-one node mapping $f \,\colon\, V_1 \to V_2$, which produces the
set of aligned node pairs $\{(v,f(v)) \mid v \in V_1\}$ (Figure
\ref{fig:main}(a)).

\vspace{0.2em}\noindent {\bf Static edge conservation.} Given an NA
between two static networks, an edge in one network is {\em conserved}
if it maps to an edge in the other network, and an edge in one network
is {\em non-conserved} if it maps to a non-adjacent node pair (i.e., a
non-edge) in the other network (Figure \ref{fig:main}(a)).  A good
static NA is a node mapping that conserves similar network
regions. That is, a good static NA should have a large number of
conserved edges and a small number of non-conserved edges.  In this
context, we measure
the quality of a static NA using the popular symmetric substructure
score (S$^3$) edge
conservation measure \citep{MAGNA}.

S$^3$ is defined as follows. Formally, the number of conserved edges is
\begin{equation*}
N_c = \sum_{(u,v) \in V_1 \otimes V_1}{\1{(u,v) \in E_1 \wedge
    (f(u),f(v)) \in E_2}},
\end{equation*}
and the number of non-conserved edges is
\begin{equation*}
\begin{split}
N_n &= \sum_{(u,v) \in V_1 \otimes V_1} \mathds{1}{[\,}
((u,v) \in E_1 \wedge (f(u),f(v)) \notin E_2) \,\,\,\vee \\[-8pt]
 & \,\,\,\,\,\qquad\qquad\qquad\,\,((u,v) \notin E_1 \wedge (f(u),f(v)) \in E_2)\,]\\
&= \abs{E_1} + \abs{E'_2} - 2N_c,
\end{split}
\end{equation*}
where $G'_2(V'_2,E'_2)$ is the subgraph of $G_2$ induced by $V'_2 =
\{f(u) \mid u \in V_1\}$, $\1{p} = 1$ if $p$ is true and $\1{p} = 0$
if $p$ is false, and $U \otimes V$ is the Cartesian product of sets
$U$ and $V$.  Then,
\begin{equation*}
\mbox{S}^3 = \frac{N_c}{N_c + N_n}.
\end{equation*}
Our implementation of S$^3$ that can compute this measure
in $O(|E_1|+|E_2|)$ time complexity is described in the Supplement.

\begin{figure}[b!]
\includegraphics[width=\linewidth]{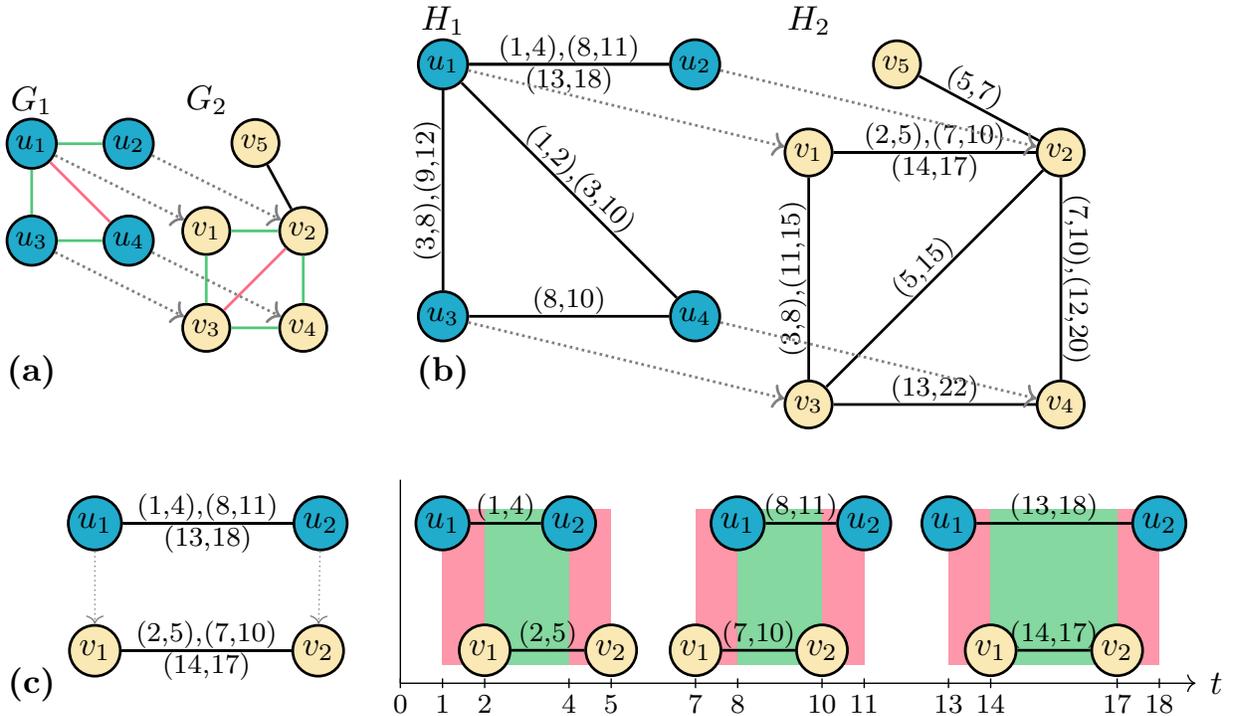}%
\caption{{\bf (a)} Two static networks $G_1(V_1,E_1)$ and
$G_2(V_2,E_2)$ (where edges between nodes in the same network are
denoted by solid lines), and a static NA between them (in this case,
$u_i$ maps to $v_i$ for $i = 1,\ldots,4$, as shown by the dotted
arrows). An edge is conserved if it maps to another edge (e.g., edge
$(u_1,u_2)$ maps to edge $(v_1,v_2)$). An edge is non-conserved if it
maps to a non-adjacent (disconnected) node pair (e.g., edge
$(u_1,u_4)$ maps to a disconnected node pair $(v_1,v_4)$). All
conserved edges are shown in green, and all non-conserved edges are
shown in red.
{\bf (b)} Two dynamic networks $H_1(V_1,T_1)$ and $H_2(V_2,T_2)$.  If
two nodes interact at least once during the network's lifetime (i.e.,
if there is at least one event between the nodes), there is a solid
line between the nodes. A given solid line can capture multiple
events. Each event is represented as $(t_s,t_e)$, where $t_s$ is its
start time, and $t_e$ is its end time.  For example, the event between
$u_3$ and $u_4$ is active from start time 8 to end time 10.  A dynamic
NA is a node mapping between the two networks (in this case, $u_i$
maps to $v_i$ for $i = 1,\ldots,4$, as shown by the dotted arrows).
}
\label{fig:main}
\end{figure}
\begin{figure}[t!]
\ContinuedFloat
\caption{
{\bf (c)} Illustration of the conserved event time (CET) and
non-conserved event time (NCET) of the mapping of node pair
$(u_1,u_2)$ to node pair $(v_1,v_2)$. On the left
are node pair $(u_1,u_2)$ from dynamic network $H_1(V_1,T_1)$ and node
pair $(v_1,v_2)$ from dynamic network $H_2(V_2,T_2)$, where
$(u_1,u_2)$ maps to $(v_1,v_2)$.
For each of the two node pairs $(u_1,u_2)$ and $(v_1,v_2)$, the event
times of the given node pair are visualized by the plot on the right, where
the given solid line in the plot indicates the start time to the end
time of the given event (i.e., the period of time during which the given node pair is
active).
Given the above, the CET between $(u_1,u_2)$ and $(v_1,v_2)$ is
the amount of time during which both $(u_1,u_2)$ and $(v_1,v_2)$ are
active at the same time (the green area).
Similarly, the NCET between
$(u_1,u_2)$ and $(v_1,v_2)$ is the amount of time during
which exactly one of $(u_1,u_2)$ or $(v_1,v_2)$
is active (the red area).
We calculate the CET between these two illustrated node pairs as follows. Since the events
$(u_1,u_2,1,4)$ and $(v_1,v_2,2,5)$ are both active from time 2 to
time 4
for a duration of $4-2 = 2$, the events $(u_1,u_2,8,11)$ and
$(v_1,v_2,7,10)$ are both active from time 8 to time 10 for a duration of
$10-8 = 2$, and the events $(u_1,u_2,13,18)$ and $(v_1,v_2,14,17)$ are
both active from time 14 to time 17 for a duration of $17-14 = 3$, the
total CET between $(u_1,u_2)$ and $(v_1,v_2)$ is $2 + 2 + 3 = 7$.
We calculate the NCET between these two illustrated node pairs as follows.
We know that $(u_1,u_2)$ is active during time periods 1 to 4, 8 to 11, and 13 to 18,
totaling a duration of $(4-1)+(11-8)+(18-13) = 11$, and that
$(v_1,v_2)$ is active during time periods 2 to 5, 7 to 10, and 14 to 17,
totaling a duration of $(5-2)+(10-7)+(17-14) = 9$.
Since NCET is the amount of time during which $(u_1,u_2)$ is active,
or $(v_1,v_2)$ is active, but not both, we need to add up the time during
which either node pair is active, and subtract the time during which
both node pairs are active (making sure to subtract twice to avoid double
counting, because of the ``but not both''
constraint). Since the time during which both node pairs are active is the
CET, the NCET  between $(u_1,u_2)$ and $(v_1,v_2)$ is
$11 + 9 - 2 \times 7 = 6$.
}
\end{figure}

\vspace{0.2em}\noindent {\bf Static node conservation.}
A good static NA should also conserve the similarity between aligned
node pairs, i.e., node conservation.  Node conservation accounts for
similarities between all pairs of nodes across the two networks.  Node
similarity can be defined in a way that depends on one's goal or
domain knowledge.  In this work, we use a node similarity measure that
is based on graphlets, as follows.

{\em Graphlets} (in the static setting) are small, connected, induced
subgraphs of a larger static network \citep{Milenkovic2008}.
Graphlets can be used to describe the extended network neighborhood of
a node in a static network via the node's graphlet degree vector
(GDV).  The {\em GDV} generalizes the degree of the node, which counts
how many edges are incident to the node, i.e., how many times the node
touches an edge (where an edge is the only graphlet on two nodes),
into the vector of graphlet degrees (i.e., GDV), which counts how many times the node
touches each of the graphlets on up to $n$ nodes, accounting in the
process for different topologically unique node symmetry groups
(automorphism orbits) that might exist within the given graphlet.  In
this work, we use all graphlets with up to four nodes, which contain 15
automorphism orbits, when calculating the GDV of a node, per
recommendations of the existing studies \citep{DynamicGraphlets,YuriyLinkPrediction}.
Hence, the GDV of a node has 15 dimensions containing counts for the
15 orbits.

Given GDVs of all nodes in two static networks $G_1(V_1,E_1)$ and
$G_2(V_2,E_2)$, where $x_u$ is the GDV of node $u$, we calculate
similarity $s(u,v)$ between nodes $u \in V_1$ and $v \in V_2$ by
relying on an existing GDV-based measure of node
similarity that was used by \citet{DynamicGraphlets}. The measure
works as follows. First, to extract GDV dimensions that contain the
most relevant information about the extended network neighborhood of
the given node, the measure first reduces dimensionality of each GDV
via principal component analysis (PCA). PCA is performed on the vector
set $\{x_w \mid w \in V_1 \cup V_2\}$, where as few as needed to
account for at least 99\% of variance in the vector set of the first
$k$ PCA components are kept. Let us denote by $y_u$ the dimensionality-reduced vector
of $x_u$ that contains the $k$ PCA components. Second, we define node
similarity $s(u,v)$ as the cosine similarity between $y_u$ and $y_v$.  Third,
given a static NA $f$, we define our node conservation measure as
$\sum_{u \in V_1}{\frac{s(u,f(u))}{\abs{V_1}}}$.

\vspace{0.2em}\noindent {\bf Objective function and optimization
  process (also known as search strategy).} MAGNA++ is a search-based
algorithm that finds a static NA by directly maximizing both
edge and node conservation. Namely, MAGNA++ maximizes the objective
function $\alpha S_E + (1-\alpha) S_N$, where $S_E$ is the
S$^3$ measure of static edge conservation described above, $S_N$ is
the graphlet-based measure of static node conservation
described above, and $\alpha$ is a parameter between 0 and 1 that controls for
the two measures. In several studies, it was shown that $\alpha$ of 0.5 yields the best results
\citep{MAGNA++,LocalVsGlobal}, which is the $\alpha$ value we use in
this study, unless otherwise noted.
Given an initial population of random static NAs, MAGNA++ evolves the
population of alignments over a number of generations while aiming to maximize
its objective function. MAGNA++ then returns the alignment from the
final generation that has the highest value of the objective
function.

\subsection{DynaMAGNA++}
\label{sec:dynamagna}

\noindent {\bf Dynamic networks.}  A {\em dynamic network} $H(V,T)$
consists of a node set $V$ and an event set $T$, where an event is a
temporal edge (Figure \ref{fig:main}(b)). An event is represented as a 4-tuple
$(u,v,t_s,t_e)$, where nodes $u$ and $v$ interact from time $t_s$ to
time $t_e$.  An event $(u,v,t_s,t_e) \in T$ is {\em active} at time
$t$ if $t_s \le t \le t_e$.  The {\em duration} of an event is the
time during which an event is active, i.e., $t_e - t_s$. There can be multiple events
between the same two nodes in the dynamic network, but no two events
between the same two nodes may be active at the same time. In
fact, if there are two events between the same two nodes that are
active at the same time, then they must be combined into a single event.

Above is the representation of a dynamic network that our study relies
on. Sometimes, dynamic data is provided in a different dynamic network
representation, most often as a discrete temporal sequence of static
network snapshots $G_1(V_1,E_1),\ldots,G_k(V_k,E_k)$. We can easily
convert the static snapshot-based representation of a dynamic network
into our event duration-based representation (i.e., into $H(V,T)$ as
defined above).  We do this as follows: if there is an edge
connecting two nodes in the $\nth{t}$ snapshot of the snapshot-based representation, then there is an event
between the two nodes that is active from time $t$ to time $t+1$ in
the event duration-based representation.
In other words, we
combine the node sets of the snapshots into a single node set $V = V_1
\cup \ldots \cup V_k$.  Then, for each snapshot $G_t$, $t=1,\ldots,k$,
we
convert each edge $(u,v) \in E_t$ into an event between nodes $u \in
V$ and $v \in V$ in the dynamic network $H(V,T)$ with start time $t$
and end time $t+1$, i.e., the event $(u,v,t,t+1)$.
This allows us to use the snapshot-based representation of a dynamic
network in our study.


\vspace{0.2em}\noindent {\bf Dynamic NA.} Given two dynamic networks $H_1(V_1,T_1)$
and $H_2(V_2,T_2)$, assuming without loss of generality that $|V_1|
\le |V_2|$, a {\em dynamic NA} between $H_1$ and $H_2$ is a one-to-one
node mapping $f \,\colon\, V_1 \to V_2$, which produces the set of
aligned node pairs $\{(v,f(v)) \mid v \in V_1\}$ (Figure
\ref{fig:main}(b)).  Note the similarity between the definitions of
static NA and dynamic NA (although the process of finding the actual
alignments is different). This makes static NA and dynamic NA fairly
comparable.

\vspace{0.2em}\noindent {\bf Dynamic edge (event) conservation.}
First, given node pair $(u_1,u_2)$ in $H_1$ that maps to node pair
$(v_1,v_2)$ in $H_2$ (Figure \ref{fig:main}(c)), we extend the notion
of a conserved or non-conserved edge from static NA to dynamic NA by
accounting for the amount of time that the mapping of $(u_1,u_2)$ to
$(v_1,v_2)$ is conserved or non-conserved (defined
below). That is, we extend the notion of a conserved or non-conserved
static edge to the amount of a conserved or non-conserved dynamic
edge (event), as follows.

Intuitively, we define the amount of a conserved event as follows.
Similar to how an edge $(u'_1,u'_2)$ in static network $G_1(V_1,E_1)$
is conserved if it maps to an edge $(v'_1,v'_2)$ in static network
$G_2(V_2,E_2)$ (and vice versa), the mapping of $(u_1,u_2)$ to $(v_1,v_2)$ is {\em
conserved} at time $t$ if both $(u_1,u_2)$ and $(v_1,v_2)$ are active
at time $t$.
We refer to the entire amount of time during which this mapping is
conserved as the {\em conserved
 event time (CET)} between $(u_1,u_2)$ and $(v_1,v_2)$. In other words,
it is the amount of
time during which both $(u_1,u_2)$ and $(v_1,v_2)$ are active at the
same time.
Formally, let
$T_{u_1u_2}$ be the set of events
between $u_1$ and $u_2$. Similarly,
let $T_{v_1v_2}$
be the set of events between $v_1$ and $v_2$.
Given this, the CET between $(u_1,u_2)$ and $(v_1,v_2)$ is
\begin{equation*}
\begin{split}
\mbox{CET}((u_1,u_2),(v_1,v_2)) = \sum_{e \in
  T_{u_1u_2}}\sum_{e' \in T_{v_1v_2}}ct(e,e'),\\
\end{split}
\end{equation*}
where the conserved time $ct(e,e') = \max(0, \min(t_e,t'_e) - \max(t_s,t'_s))$ is
the amount of time during which events $e = (u_1,u_2,t_s,t_e)$ and $e'
= (v_1,v_2,t'_s,t'_e)$ are active at the same time, i.e., $ct(e,e')$
is the length of the overlap of the intervals $[t_s,t_e]$ and $[t'_s,t'_e]$.

Intuitively, we define the amount of a non-conserved event as follows.
Similar to how an edge $(u'_1,u'_2)$ in $G_1$ is non-conserved if
it maps to a disconnected node pair $(v'_1,v'_2)$ in $G_2$ (or vice versa),
the mapping of $(u_1,u_2)$ to $(v_1,v_2)$ is
{\em non-conserved} at time $t$ if exactly one of $(u_1,u_2)$ or
$(v_1,v_2)$ is active at time $t$.
We refer to the entire amount of time during which this mapping is
non-conserved as
the {\em non-conserved event time (NCET)} between $(u_1,u_2)$ and
$(v_1,v_2)$. In other words, it is the amount of time during which
$(u_1,u_2)$ is active, or $(v_1,v_2)$ is active, but not both are active at the
same time.
Formally, the NCET between $(u_1,u_2)$ and $(v_1,v_2)$ is
\begin{equation*}
\mbox{NCET}((u_1,u_2),(v_1,v_2))
 = \sum_{e \in T_{u_1u_2}} d(e) + \sum_{e' \in
  T_{v_1v_2}} d(e') - 2\sum_{e \in T_{u_1u_2}}\sum_{e' \in T_{v_1v_2}} ct(e,e'),
\end{equation*}
where $d(e)$ is the duration of event $e$, i.e., the amount of time during
which $e$ is active. We make sure to subtract twice the amount of time
during which $(u_1,u_2)$ and $(v_1,v_2)$ are both active due to the
above ``but not both are active at the same time'' constraint.

Second, given the above definitions of CET and NCET between two node pairs
$(u_1,u_2)$ and $(v_1,v_2)$,
we extend the S$^3$ measure of static node conservation to a new dynamic
S$^3$ (DS$^3$) measure of dynamic edge (event) conservation, which we
propose as a contribution of this study. To define DS$^3$, we need to
introduce the notion of
CET between all node pairs across the entire alignment (rather than
between just two aligned node pairs), henceforth simply referred to as
\emph{alignment CET}, which is the sum of CET between
all node pair mappings between $H_1$ and $H_2$.
Analogously, we need to define the notion of \emph{alignment NCET},
which is the sum of NCET between all node pair mappings between $H_1$ and
$H_2$. Alignment CET measures the amount of event conservation of the entire
alignment and alignment NCET measures the amount of event non-conservation
of the entire alignment.  A good dynamic NA is a node mapping that
conserves similar evolving network regions. That is, a good dynamic NA
should have high alignment CET and low alignment NCET, which is what
DS$^3$ aims to capture. Formally, alignment CET is
\begin{equation*}
T_c = \sum_{(u,v) \in V_1 \otimes V_1} \mbox{CET}(\,(u,v),\,\,(f(u),f(v))\,)
\end{equation*}
and alignment NCET is
\begin{equation*}
T_n = \sum_{(u,v) \in V_1 \otimes V_1} \mbox{NCET}(\,(u,v),\,\,(f(u),f(v))\,).
\end{equation*}
Then,
\begin{equation*}
\mbox{DS}^3 = \frac{T_c}{T_c + T_n}.
\end{equation*}
Our implementation of
DS$^3$ that can compute this measure in $O(|T_1|+|T_2|)$ time
complexity is described in the Supplement.

We note that there are many real-world networks that contain events
with durations that are significantly less than the entire time window
of the network, called ``bursty'' events.  Examples of networks
containing bursty events are e-mail communication networks, economic
networks that model transactions, and brain networks constructed from
oxygen level correlations as measured by fMRI scanning, each of whose
events last much less than a second while the networks' time windows
span minutes to hours \citep{ModernTemporal}.  Since bursty events are
so short, small perturbations in the event times can greatly affect
the resulting dynamic edge (event) conservation value. Thus, in order
to allow our DS$^3$ measure to be more robust to perturbations in the
event times, one may simply extend the duration of each event in the
network by some time $\Delta t$.
Extending the duration of each event by $\Delta t$ will account for
perturbations in event times of up to $\frac{\Delta t}{2}$ due to the
following.  Given two events $(u_1,u_2,t,t)$ and $(v_1,v_2,t',t')$
with durations of 0, where $t' = t + \frac{\Delta t}{2}$, the
conserved time $ct(\cdot,\cdot)$ between the two events is 0.  Thus,
if we want to consider the two events as conserved, we can increase the
durations of both events by $\Delta t$ to create the modified events
$(u_1,u_2,t,t+\Delta t)$ and $(v_1,v_2,t',t'+\Delta t)$, which results
in a conserved time of $\frac{\Delta t}{2}$ for the two modified
events.  While we do not use this technique in our work since we do
not use networks with bursty events,
others might in the future, and if so, this needs to be considered
when performing dynamic NA.

\vspace{0.2em}\noindent {\bf Dynamic node conservation.}
Just as for static NA, a good dynamic NA method should also conserve the
similarity between aligned node pairs, i.e., node conservation.
To take advantage of the temporal  information encoded in
dynamic networks that are being aligned and also to make dynamic NA
as fairly comparable as possible to static NA, in this work, we rely on a
measure of node similarity based on dynamic graphlets, as follows.

Dynamic graphlets are an extension of static graphlets (Section
\ref{sec:magna}) from the static setting to the dynamic setting by
accounting for temporal information in the dynamic network.  While
static graphlets can be used to capture the static extended network
neighborhood of a node, dynamic graphlets can be used to capture how
the extended neighborhood of a node changes over time.  To describe
dynamic graphlets formally, we first present the notion of a $\Delta
t$-time-respecting path and a $\Delta t$-connected network. A $\Delta
t$-time-respecting path is a sequence of events that connect two nodes
such that for any two consecutive events in the sequence, the end time
of the earlier event and the start time of the later event are
within $\Delta t$ time of each other (i.e., are $\Delta t$-adjacent).
A dynamic network
is $\Delta t$-connected if for each pair of nodes in the network,
there is a $\Delta t$-time-respecting path between the two
nodes. Then, just as a static graphlet is an equivalence class of
isomorphic connected subgraphs (Section \ref{sec:magna}), a {\em dynamic graphlet}
is an equivalence class of isomorphic
$\Delta t$-connected dynamic subgraphs, where two graphlets are
equivalent if they both have the same {\em relative} temporal order of
events. We use $\Delta t = 1$ in this work, per recommendations by
\citet{DynamicGraphlets}. Just as the GDV of a node in a static
network is a
topological descriptor for the extended neighborhood the node, there
exists the {\em dynamic GDV (DGDV)} of a node in a dynamic network,
which describes how the extended neighborhood of a node changes over
time.  Specifically, just as the GDV of a node
counts how many times the node touches each static graphlet at each of
its automorphism orbits, the DGDV of a node counts how many times the
node touches each dynamic graphlet at each of its orbits. Dynamic
graphlets have a similar notion of
orbits as static graphlets do, which now depend on both topological and
temporal positions of a node within the dynamic graphlet.  To make
things comparable as fairly as possible to static NA, and per recommendations by
\citet{DynamicGraphlets}, in this work,
we use dynamic graphlets with up to four nodes and six events, which
contain 3,727 automorphism orbits, when
calculating the DGDV of a node. Hence, the DGDV of a node has 3,727
dimensions containing counts for the 3,727 orbits.

Given the DGDVs of all nodes in two dynamic networks $H_1(V_1,T_1)$ and
$H_2(V_2,T_2)$, we calculate similarity $s(u,v)$ between nodes $u \in V_1$
and $v \in V_2$, in the same way as in Section \ref{sec:magna} (by relying on the
PCA-based dimensionality reduction of all nodes' DGVDs, computing
cosine similarity between the dimensionality-reduced PCA vectors, and
accounting for resulting cosine similarities between all pairs of
nodes across the compared networks to obtain the total dynamic node
conservation).

\vspace{0.2em}\noindent {\bf Objective function and optimization
  process (also known as search strategy).}  DynaMAGNA++ is a search-based
algorithm that finds a dynamic NA by directly maximizing both dynamic
edge (event) and node
conservation. Namely, DynaMAGNA++ maximizes the objective function $\alpha S_T +
(1-\alpha) S_N$, where $S_T$ is the DS$^3$ measure of dynamic edge
conservation described above, $S_N$ is the DGDV-based measure of
dynamic node conservation described above, and $\alpha$ is a parameter
between 0 and 1 that controls for the two measures.
To make DynaMAGNA++ fairly comparable to MAGNA++, here we also use
MAGNA++'s best $\alpha$ value of 0.5, unless otherwise noted.
Given an initial population of random dynamic NAs, DynaMAGNA++ evolves the
population of alignments over a number of generations while aiming to maximize
its objective function. DynaMAGNA++ then returns the alignment from
the final generation that has the highest value of the objective
function.

\vspace{0.2em}\noindent {\bf Time complexity.}
To align two dynamic networks $H_1(V_1,T_1)$ and $H_2(V_2,T_2)$,
DynaMAGNA++ evolves a population of $p$ alignments over $N$
generations.
It does so by using its crossover function (see \citet{MAGNA} for
details) to combine pairs of parent alignments in the given population
into child alignments, for each generation. For each generation, the dynamic edge (event)
conservation, dynamic node conservation, and crossover of $O(p)$
alignments are calculated.  Since dynamic edge conservation takes
$O(|T_1|+|T_2|)$ to compute, dynamic node conservation takes
$O(|V_1|)$ time to compute, crossover takes $O(|V_2|)$ time to
compute, and $\abs{V_1} \le \abs{V_2}$, the time complexity of
DynaMAGNA++ is $O(N p |V_2| + N p (|T_1|+|T_2|))$. Note that the
calculation of dynamic edge and dynamic node conservation in
DynaMAGNA++ is parallelized. This allows DynaMAGNA++ to be run on
multiple cores, which empirically results in close to liner speedup.

\vspace{0.2em}\noindent {\bf Other parameters.}
Given an initial population of dynamic NAs, DynaMAGNA++ evolves the
population for up to a specified number of generations or until it reaches
a stopping criterion. For each generation, DynaMAGNA++ keeps an elite
fraction of alignments from the current generation's population for the
next generation's population.
In addition to the dynamic edge and node conservation measures, and the $\alpha$ parameter
that controls for the contribution of the two measures, the remaining
parameters of DynaMAGNA++ are (i) the initial
population, (ii) the size of the population, (iii), the
maximum number of generations, (iv) the elite fraction, and (v) the stopping criterion.
For DynaMAGNA++, we use a  population of 15,000 alignments initialized
randomly, as in the original MAGNA++ paper, unless otherwise noted. We
specify a maximum of 10,000
generations, since the alignments that we test all converge by
10,000 generations.
The elite fraction is 0.5, as in the original MAGNA++ paper. The algorithm stops when the highest
objective function value in the population has increased less than
0.0001 within the last 500 generations, since the alignments that we
test do not increase by a significant amount after this point.

To fairly compare DynaMAGNA++ against MAGNA++, we aim to set the
parameters of both methods to be as similar as possible.  So, other
than MAGNA++'s edge and node conservation measures,
the remaining parameters of MAGNA++ are the same as for DynaMAGNA++.
This way, any differences that we see between results of DynaMAGNA++
and results of MAGNA++ will be the consequence of the differences of
the two methods' edge and node conservation measures, i.e., of
accounting for temporal information in the network with DynaMAGNA++
and ignoring this information with MAGNA++. In other words, any
differences that we see between results of DynaMAGNA++ and results of
MAGNA++ will fairly reflect differences between dynamic NA and static
NA.

\section{Results and discussion}
\label{sec:results}

Since there are no other dynamic NA methods to compare against, we
compare DynaMAGNA++ to the next best option, namely its static NA
counterpart.  That is, we compare DynaMAGNA++ when it is used to align
two dynamic networks, to MAGNA++ when it is used to align static
versions of the two dynamic networks. By ``static versions'', we mean
that we ``flatten'' or ``aggregate'' a dynamic network into a static
network that will have the same set of nodes as the dynamic network
and a static edge will exist between two nodes in the static network
if there is at least one event between the same two nodes in the
dynamic network.  This network aggregation simulates the common
practice where network analysis of time-evolving systems is done in a
static manner, by ignoring their temporal information
\citep{DynamicGraphlets,ModernTemporal}.

We evaluate DynaMAGNA++ and MAGNA++ on synthetic and real-world
dynamic networks, as described in the following sections.
Note that there is no need to compare DynaMAGNA++ to any other static
NA method besides MAGNA++, because MAGNA++ was recently shown in a
systematic and comprehensive manner to be superior to seven other
state-of-the-art static NA methods (Section \ref{sec:intro}). So, by transitivity, to demonstrate
that dynamic NA is superior to static NA, it is sufficient to
demonstrate that DynaMAGNA++ is superior to MAGNA++.

\subsection{Evaluation using synthetic networks}
\label{sec:syntheticmain}

\vspace{0.2em}\noindent\textbf{Motivation.} A good NA approach should
be able to produce high-quality alignments
between networks that are similar and low-quality alignments between
networks that are dissimilar \citep{AlignmentFree}.  In this test on
synthetic networks, ``similar'' means networks that originate from the
same network model, and ``dissimilar'' means networks that originate
from different network models. So, we refer to this test as network
discrimination.  Thus, in this section, we evaluate the network discrimination
performance of DynaMAGNA++ and MAGNA++.


\vspace{0.2em}\noindent\textbf{Data.} We perform this evaluation on a set of biologically inspired synthetic
networks.  Specifically, we generate 20 dynamic networks using four
biologically inspired network evolution models (or versions of the
same model with different parameter values) that simulate the
evolution of PINs, resulting in five networks per model
\citep{DynamicGraphlets}.  The four models we use are (i) GEO-GD with
$p = 0.3$, (ii) GEO-GD with $p = 0.7$, (iii) SF-GD with $p = 0.3$ and
$q = 0.7$, and (iv) SF-GD with $p = 0.7$ and $q = 0.6$, where GEO-GD
is a geometric gene duplication model with probability
cut-off and SF-GD is a scale-free gene duplication model
\citep{BioNetworkModel}.  \citet{DynamicGraphlets} generalized the
static versions of these models to their dynamic counterparts, and we
rely on the same model networks as those used by
\citet{DynamicGraphlets} (see their paper for details). The synthetic
networks are represented as snapshots, with 1,000 nodes in each of the
networks, an average of 24 snapshots per network, and an average of 162 edges per
snapshot, where any variation in network sizes is caused by the
different parameter values of the considered network models.

\begin{figure}[t!]
\centering
\includegraphics[width=0.85\linewidth]{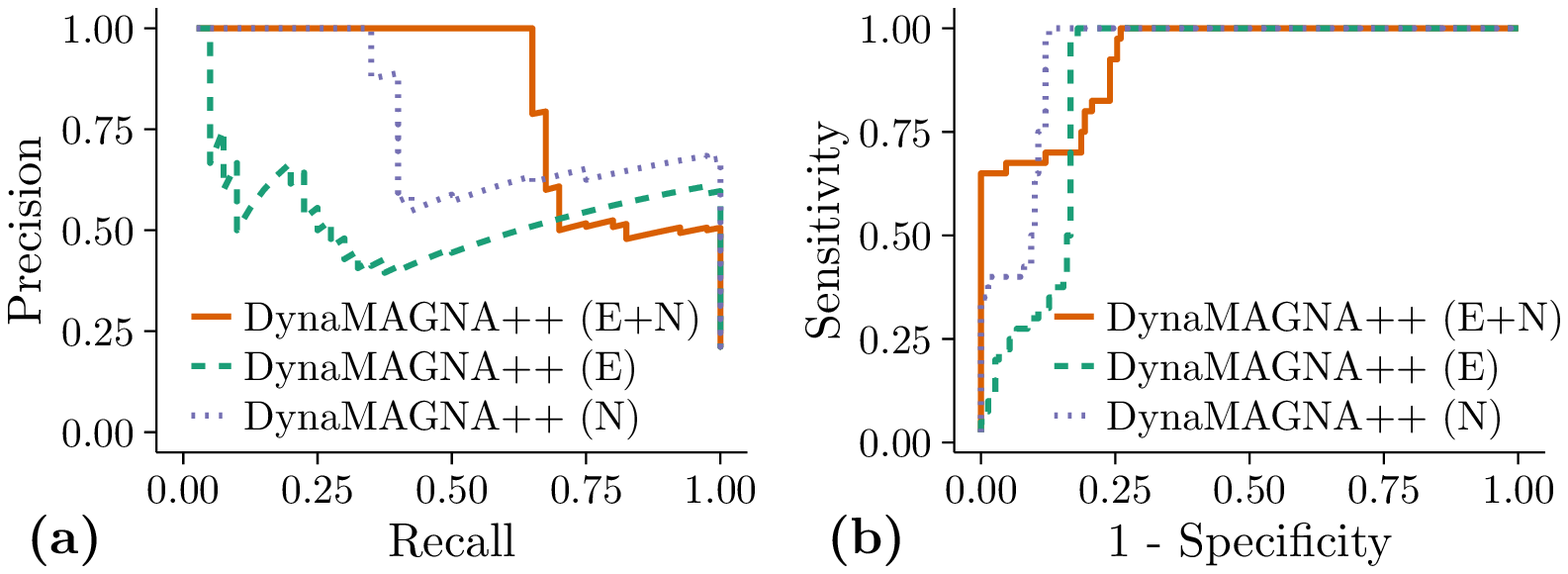}%
    \caption{Network discrimination performance of
DynaMAGNA++, while optimizing both dynamic edge and node
conservation (E+N), dynamic edge conservation alone (E), and
dynamic node conservation alone (N), for biological synthetic networks, with respect
to {\bf (a)} precision-recall curve and {\bf (b)} ROC curve.}
\label{fig:test1a}
\end{figure}

\begin{table}[t!]
\centering
\begin{tabular}{rcccc}
  \toprule
\input{exp_1k_a_objective_function_algorithm_comparison_ss}
\end{tabular}
\caption{Network discrimination performance of DynaMAGNA++, while
optimizing both dynamic edge and node conservation (E+N), dynamic edge
conservation alone (E), and dynamic node conservation alone (N), for
biological synthetic networks, with respect to the area under the
precision-recall curve (AUPR), F-score at which precision and recall
cross and are thus equal (F-score$_{\mbox{cross}}$), maximum F-score
(F-score$_{\mbox{max}}$), and the area under the ROC curve (AUROC).}
\label{tab:test1a}
\end{table}

\vspace{0.2em}\noindent\textbf{Evaluation measures.} We calculate
network discrimination performance of both methods as follows.  We
align all $20 \choose 2$ pairs of the synthetic networks using
DynaMAGNA++ and MAGNA++.  The higher the alignment quality between
pairs of similar networks (i.e., networks coming from the same model)
and the lower the alignment between pairs of dissimilar networks
(i.e., networks coming from different models), the better the NA
method. Here, by alignment quality between two networks that the
given method (DynaMAGNA++ or MAGNA++) identifies, we mean the method's
objective function value for the alignment of the two networks that is
returned by the method (Section \ref{sec:methods}).  For a given
method, given all $20 \choose 2$ alignment quality values, we
summarize the method's network discrimination performance using
precision-recall and receiver operating characteristic (ROC)
frameworks. Specifically, given the alignment quality values of all
pairs of networks, for some given threshold $r$, a good NA method
should result in alignment quality greater than $r$ for pairs of
similar networks and in alignment quality smaller than $r$ for pairs
of dissimilar networks.  So, for a given threshold $r$, we compute
accuracy in terms of precision, the fraction of network pairs that are
similar and with alignment quality greater than $r$ out of all network
pairs with alignment quality greater than $r$, and recall, the
fraction of network pairs that are similar and with alignment quality
greater than $r$ out of all similar network pairs.  Varying the
threshold $r$ for all $r \ge 0$ (i.e., for $r$ between 0 and the
maximum observed alignment quality value, in increments of the smallest
difference between any pair of observed alignment quality values)
while plotting the resulting precision and recall values on the the
$y$ and $x$ axes, respectively, gives us the precision-recall curve.
Then, we compute the area under the precision-recall curve (AUPR), the
F-score (harmonic mean of precision and recall) at which precision and
recall cross and are thus equal (F-score$_{\mbox{cross}}$), and the
maximum F-score over all threshold $r$ values
(F-score$_{\mbox{max}}$).  For a given threshold $r$, we also compute
method accuracy in terms of sensitivity, which is the same as recall,
and specificity, the fraction of network pairs that are dissimilar and
with alignment quality less than $r$ out of all network pairs that are
dissimilar.  Varying the threshold $r$ for all $r \ge 0$ while
plotting the resulting sensitivity and $1 - $ specificity values on
the the $y$ and $x$ axes, respectively, gives us the receiver
operating characteristic (ROC) curve.  Then, we compute the area under
the ROC curve (AUROC).

\vspace{0.2em}\noindent\textbf{Results.} First, we aim to test whether
optimizing both dynamic edge (event) conservation and dynamic node conservation in
DynaMAGNA++ is better than optimizing either dynamic edge conservation
alone or dynamic node conservation alone, since it was shown for
MAGNA++ in previous studies that optimizing both static edge conservation and
static node conservation performs better than optimizing either static
edge conservation alone or static node conservation alone
\citep{MAGNA++,LocalVsGlobal}.  So, we compare three different
versions of DynaMAGNA++ that differ in their optimization
functions. Namely, the three versions optimize: (i) a combination of
dynamic edge conservation and dynamic node conservation (corresponding
to $\alpha = 0.5$, named DynaMAGNA++ (E+N)), (ii) dynamic edge
conservation only (corresponding to $\alpha = 1$, named DynaMAGNA++
(E)), and (iii) dynamic node conservation only (corresponding to
$\alpha = 0$, named DynaMAGNA++ (N)) (Section \ref{sec:methods}).  We
find that overall DynaMAGNA++ (E+N) performs better than DynaMAGNA++
(E) and DynaMAGNA++ (N) (Figure \ref{fig:test1a} and Table
\ref{tab:test1a}), especially in terms of AUPR (as well as
F-score$_{\mbox{cross}}$), which is more credible than AUROC when there is an
imbalance between the number of similar pairs and dissimilar pairs, as
is the case in our study.  So, henceforth, in the main paper, we only
report results for DynaMAGNA++ (E+N) and refer to it simply as
DynaMAGNA++. We report corresponding results for DynaMAGNA++ (E) and
DynaMAGNA++ (N) in the Supplement.  Thus, we fairly compare DynaMAGNA++ and
MAGNA++, both using the $\alpha$ parameter value of 0.5.

\begin{figure}[t!]
\centering
\includegraphics[width=0.85\linewidth]{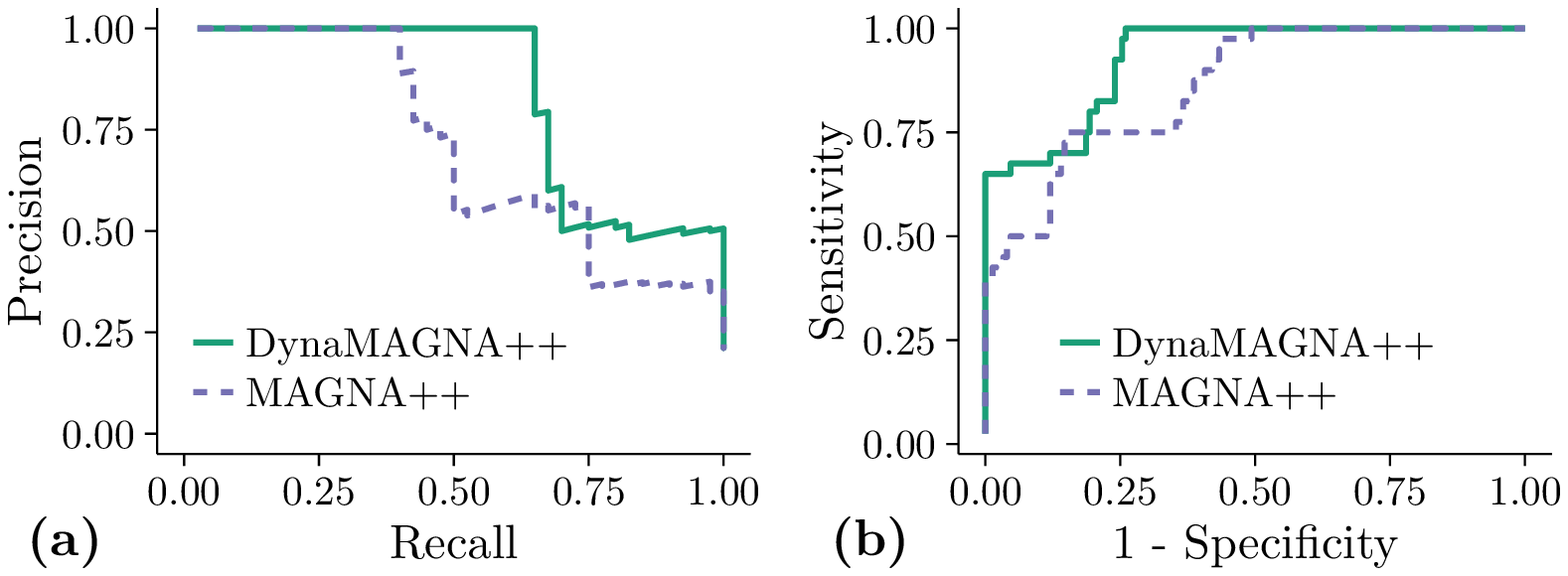}%
    \caption{Network discrimination performance of
DynaMAGNA++ and MAGNA++ for biological synthetic networks with respect
to {\bf (a)} precision-recall curve and {\bf (b)} ROC curve.}
\label{fig:test1b}
\end{figure}

\begin{table}[t!]
\centering
\begin{tabular}{rcccc}
  \toprule
\input{exp_1k_s_objective_function_algorithm_comparison_ss}
\end{tabular}
\caption{Network discrimination performance of DynaMAGNA++ and MAGNA++,
for biological synthetic networks, with respect to the same measures
as in Table \ref{tab:test1a}.}
\label{tab:test1b}
\end{table}

Second, and most importantly, we aim to answer whether dynamic NA is better than static NA,
by comparing the network discrimination performance of DynaMAGNA++
and MAGNA++.  We find that
DynaMAGNA++ performs better than
MAGNA++ in the task of network discrimination
with respect to all considered measures of NA quality (Figure
\ref{fig:test1b} and Table \ref{tab:test1b}; also, see the Supplement).

To illustrate generalizability of dynamic NA to other domains, we also
evaluate the performance of DynaMAGNA++ and MAGNA++ on synthetic
networks under a social network evolution model. We
find that similar results (superiority of DynaMAGNA++ over MAGNA++)
hold for the synthetic social networks as well
(see the Supplement).

In summary, under fair comparison conditions, we demonstrate that
 dynamic NA is superior to static NA for synthetic dynamic networks.

\subsection{Evaluation using real-world networks}
\label{sec:realworldmain}

\vspace{0.2em}\noindent {\bf Motivation.}
Here, we still evaluate
whether the given method produces high-quality
alignments for similar networks and low-quality alignments for
dissimilar networks. However, here, for real-world networks, the
notion of similarity that we use is different than for the synthetic
networks above, because for real-world networks, we do not know which
network models they belong to \citep{AlignmentFree}. Specifically,
here, we align an original real-world network to randomized (noisy)
versions of the original network (see below), where we
vary the noise level.  The larger the noise level, the more dissimilar
the aligned networks are, and consequently, the lower the alignment
quality should be.

\vspace{0.2em}\noindent {\bf Zebra network.}
Since there is a lack of available dynamic {\em molecular} networks
(Section \ref{sec:intro}), we
begin our evaluation of real-world networks on an alternative
biological network type, namely an ecological network.
The original real-world network that we use is the Grevy's zebra
proximity network \citep{Zebra}, which contains information on interactions
between 27 zebras in Kenya over 58 days. The data was collected by
driving a predetermined route each day while
searching for herds. There are 779 events in the network.
We also report results for another animal proximity network that contains
information on interactions between 28 onagers, a species
that is closely related to the Grevy's zebra, mostly in the Supplement. The
onager proximity network contains 28 nodes and 522 events.




Since the difference between dynamic NA and static NA is
that the former accounts for the temporal aspect of the data more
explicitly than the latter, to properly validate results for dynamic
NA, as strict randomization
scheme as possible should be used when creating randomized (noisy)
versions of the
original dynamic network that will be aligned to the original
network. By ``as strict as possible'', we mean that we want to use a
randomization scheme that preserves as much structure (i.e., topology)
as possible of
the dynamic network and randomizes only the temporal aspect of the
network. This way, the only difference observed between DynaMAGNA++'s
and MAGNA++'s performance will be the consequence of considering the
temporal aspect of the data. For this reason, we randomize the
original network using the following model per recommendations by
\citet{ModernTemporal}.
In order to randomize the original dynamic network $H(V,T)$ to a noise level, first, we arbitrarily number all $m$ events in
the network as $T = \{e_1,e_2,\ldots,e_m\}$. Then, for each event
$e_i$, with probability $p$ (where $p$ is the noise level) we randomly
select another event $e_j, j \ne i$, and swap the time stamps of the
two events.  Since we only swap the time stamps, this randomization
scheme conserves the total number of events and the
structure of the flattened version of the original dynamic network.
We study 10 different noise levels (from 0\% to 100\% in smaller
increments initially and larger increments toward the end). For each
noise level, we generate five randomized versions of the original
network and report results averaged over the five randomization runs.

\begin{figure}
\centering
\includegraphics[width=0.85\linewidth]{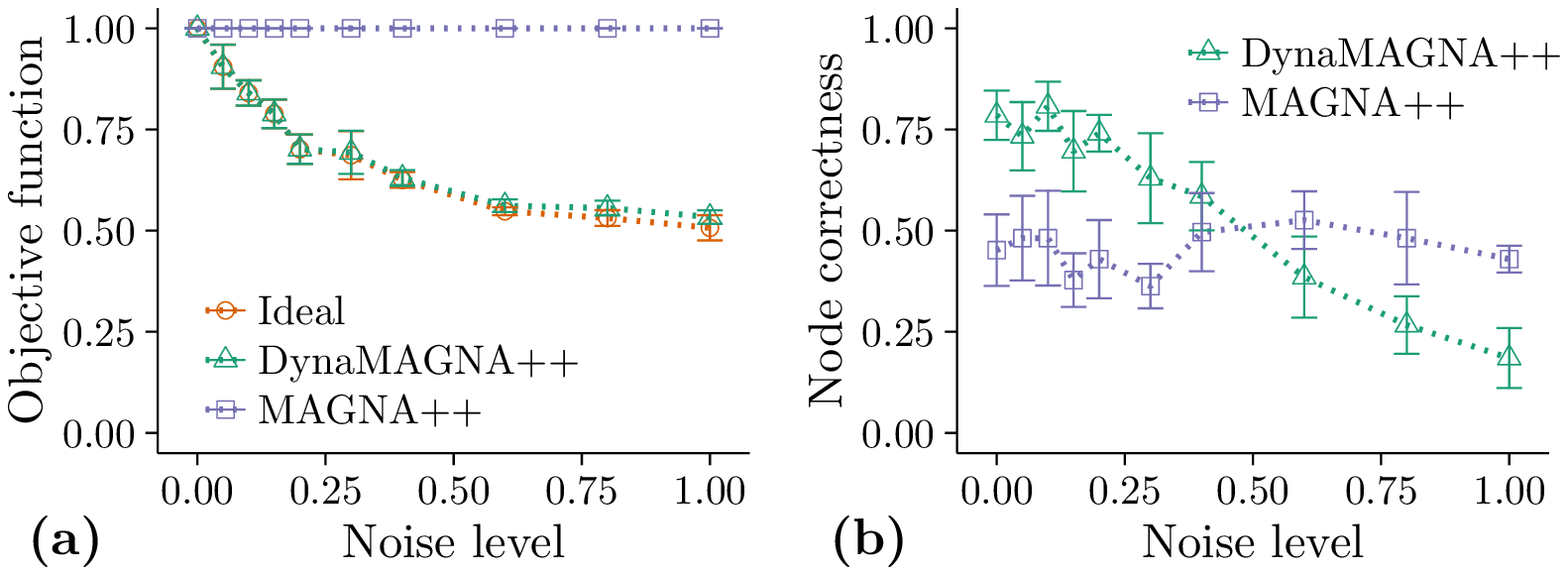}%
    \caption{Alignment quality of DynaMAGNA++ and MAGNA++ as a
function of noise level when aligning the original Grevy's zebra
network to randomized (noisy) versions of the original network. In
this figure, the randomization is as strict as possible, as it
conserves all structure of the flattened version of the original
dynamic network and only randomly ``shuffles'' the given percentage
(noise level) of its event time stamps.  Alignment quality is shown
with respect to {\bf (a)} each method's objective function, and {\bf
(b)} node correctness.  ``Ideal'' in panel (a) shows the quality of
perfect alignments, with respect to DynaMAGNA++'s objective function.}
\label{fig:test2a}
\end{figure}

We evaluate DynaMAGNA++ and MAGNA++'s performance as follows. First,
for a good method, alignment quality should
decrease as the noise level increases, since the original network and
its randomized version become more dissimilar with this increase.  As
in Section \ref{sec:syntheticmain}, one measure of alignment
quality that we use is each method's objective function.
Another measure that we use is
node correctness.  Node correctness of an alignment is the fraction of
correctly aligned node pairs (according to the ground truth node
mapping) out of all aligned node pairs.  Given
that our original network and its randomized versions have the same
set of nodes, we know which nodes in the original network correspond
to which nodes in the given randomized network. That is, we know the
ground truth mapping between the aligned networks (which henceforth we
refer to as the perfect mapping). Hence, we can measure node
correctness between the networks.
Thus, we evaluate each method's alignment quality using the method's
objective function as well as node correctness, with the expectation
that for a good method, alignment quality should decrease with
increase in the noise level.

Second, since we
know the perfect alignment between the original network and each of
its randomized versions, we compute the
``ideal'' alignment quality, i.e., the quality of the perfect
alignment, as measured by DynaMAGNA++'s objective function.
Here, the expectation is that a good method's alignment quality should
mimic well the quality of the perfect alignment.

Third, we expect DynaMAGNA++'s alignment quality to be superior to
MAGNA++'s alignment quality with respect to node correctness for lower
(meaningful) noise levels, if it is indeed true that dynamic NA
is superior to static NA. We do not expect this superiority
for higher noise levels, since at such noise levels, networks being
aligned are highly randomized and thus a good method should produce
low-quality alignments.

Indeed, our results confirm all three of the above expectations
(Figure \ref{fig:test2a}; also, see the Supplement). Specifically, first, DynaMAGNA++'s alignment quality
indeed decreases with the increase in the noise level with respect to
both its objective function (Figure \ref{fig:test2a}(a)) as well as
node correctness (Figure \ref{fig:test2a}(b)).  On the other hand,
MAGNA++'s alignment quality
stays constant with increase in the noise level. In
other words, MAGNA++ produces alignments of the same quality for low noise
levels (where network structure is meaningful) as it does for high
noise levels (where network structure is random).  Second,
DynaMAGNA++'s alignment quality follows closely
the quality of the perfect alignments
(while MAGNA++ does not) (Figure \ref{fig:test2a}(a)).
Third, DynaMAGNA++ achieves higher node correctness than MAGNA++ at
lower (meaningful) noise levels. This is not a
surprise, since DynaMAGNA++ explicitly uses the temporal information
in the aligned networks, while MAGNA++ does not.  Thus, in summary,
dynamic NA outperforms static
NA.  We observe similar results for the onager network (see the Supplement).

\begin{figure}
\centering
\includegraphics[width=0.85\linewidth]{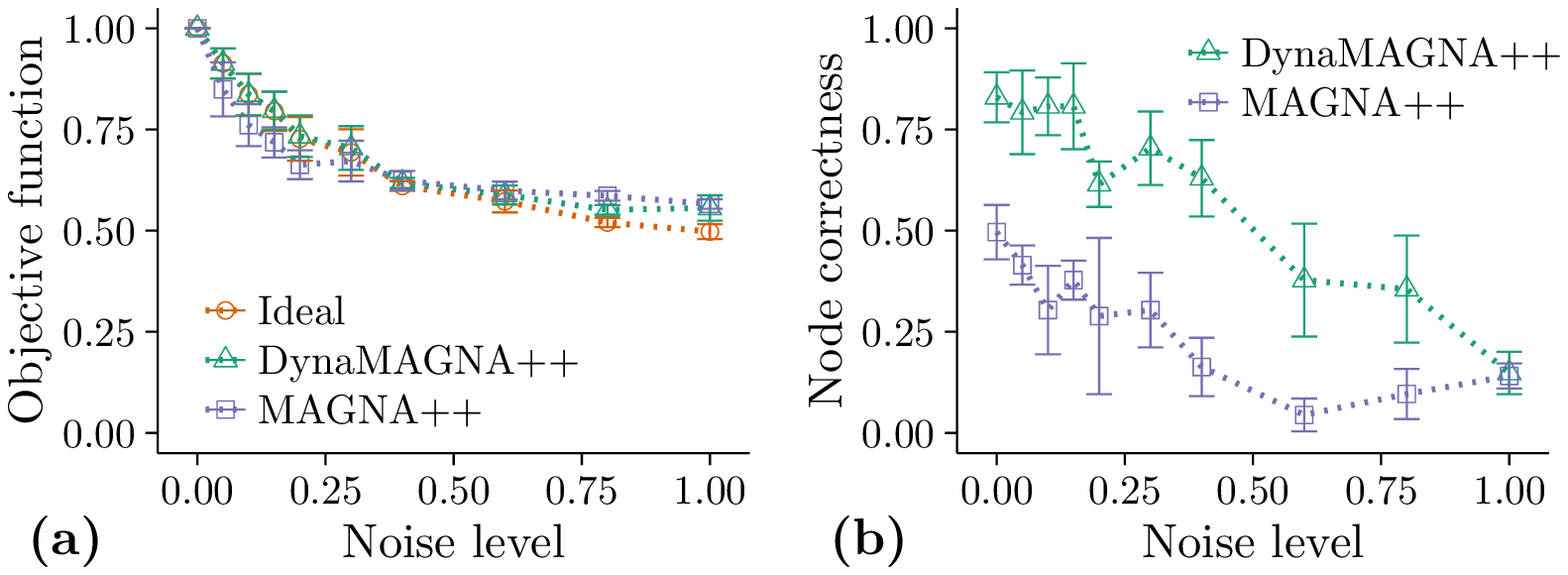}%
    \caption{Alignment quality of DynaMAGNA++ and MAGNA++
for the Grevy's zebra network. The figure can be interpreted in the
same way as Figure \ref{fig:test2a}, except that here, the randomization used to
create the noisy networks does not conserve the structure
of the flattened version of the original dynamic network.}
\label{fig:test2b}
\end{figure}

The above complete failure of MAGNA++ to produce alignments of
decreasing quality as the noise level increases is due to the strict
randomization scheme that we use to create the noisy versions of the
original network, which conserves all structure of the
flattened version of the original dynamic network. Recall that we use
the strict randomization scheme to ensure that the results of
DynaMAGNA++ are meaningful. Yet, to give as fair
advantage as possible to static NA, we produce a different set of
noisy versions of the original network using somewhat more flexible
randomization scheme that does not conserve the structure
of the flattened version of the original dynamic network, per
recommendations of \citet{ModernTemporal}. This
randomization scheme works as follows.
In order to randomize the original dynamic network $H(V,T)$ to a
certain noise level, first, we arbitrarily number all $m$ events in
the network as $T = \{e_1,e_2,\ldots,e_m\}$.
Then, for each event $e_i$, with
probability $p$ (where $p$ is the noise level) we randomly select
an event $e_{i'}$, and we rewire the two events. That is, given
$e_i = (u,v,t_s,t_e)$ and $e_{i'} = (u',v',t'_s,t'_e)$, we either set
$e_i = (u,v',t_s,t_e)$ and $e_{i'} = (u',v,t_s,t_e)$ with probability
0.5, or we set $e_i =
(u,u',t_s,t_e)$ and $e_{i'} = (v,v',t_s,t_e)$ with probability 0.5.
If the rewiring creates a loop (i.e., an event from a node to itself) or multiple
link (i.e., duplicate events between the same nodes),
then we undo it and randomly select another event $e_{i'}$.  This
randomization scheme conserves the entire set of time stamps of the original
network, but it does not preserve the structure of the
flattened network.  We study 10 different noise levels (from 0\% to 100\% in smaller
increments initially and larger increments toward the end). For each noise
level, we generate five randomized versions of the original network
and report results averaged over the five randomization runs.
While now MAGNA++'s alignment quality also decreases with increase
in the noise level and also MAGNA++ closely follows the quality of the perfect alignments,
as it should (Figure \ref{fig:test2b}(a); also, see the Supplement), DynaMAGNA++ is still superior to
MAGNA++ with respect to node correctness (Figure \ref{fig:test2b}(b)),
which again implies that dynamic NA is superior to static NA.

Because the results are consistent independent of the randomization
scheme that is used to produce noisy networks, and since the strict
scheme should be used to correctly evaluate DynaMAGNA++'s correctness,
henceforth, we report results only for the strict randomization
scheme.

\begin{figure}
\centering
\includegraphics[width=0.85\linewidth]{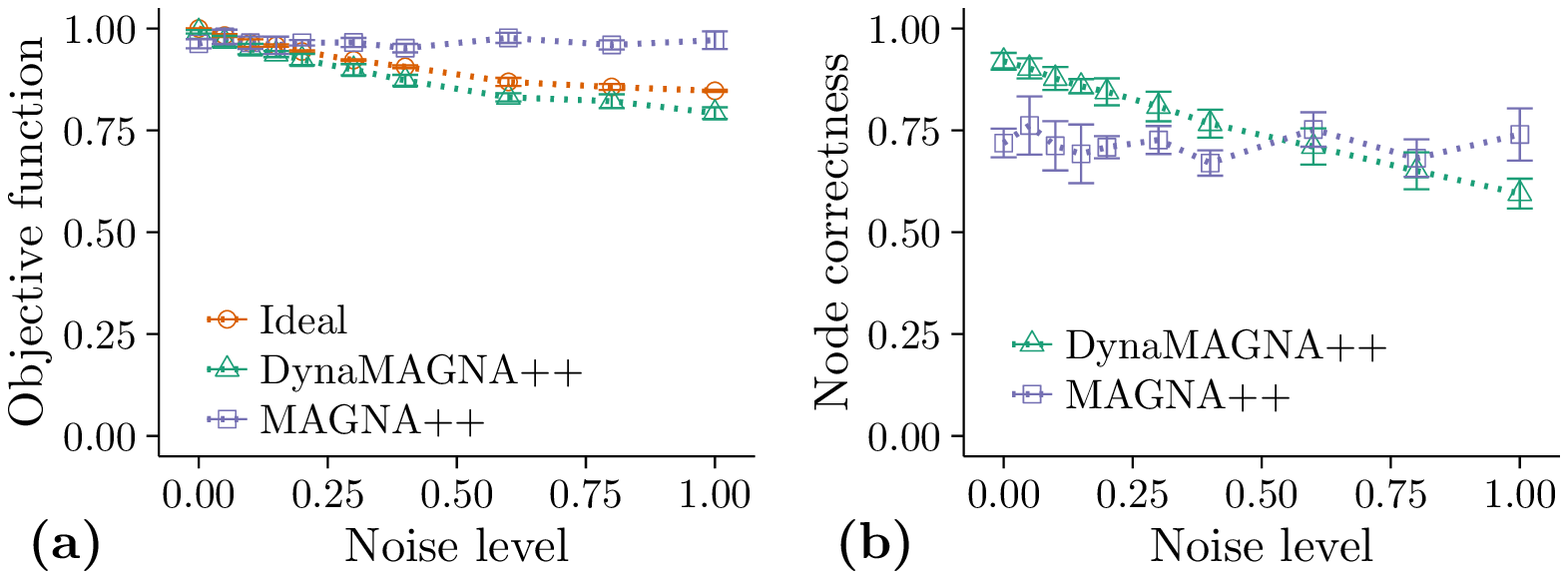}%
    \caption{Alignment quality of DynaMAGNA++ and MAGNA++ for the
      yeast network. The figure can be interpreted in the same way as
      Figure \ref{fig:test2a}.}
\label{fig:test2b}
\end{figure}

\vspace{0.2em}\noindent {\bf Yeast network.}
Since there is a lack of available \emph{experimental} dynamic molecular networks, we
continue our evaluation of real-world
networks on the next best available dynamic molecular network
option. Namely, we create a dynamic yeast PIN from an artificial
temporal sequence of static yeast PINs. Here, the static
PINs
that are used as snapshots of the dynamic PIN are
 are all real-world networks, it is just their temporal sequence
that is artificial. The sequence consists of six static PIN snapshots: a
high-confidence {\em S. cerevisiae} (yeast) PIN with 1,004
proteins and 8,323 interactions,
and five lower-confidence yeast PINs
constructed by adding to the high-confidence PIN 5\%, 10\%, 15\%,
20\%, or 25\% of lower-confidence interactions; the higher-scoring
lower-confidence interactions are added first. Clearly, the five
lower-confidence PINs have the same 1,004 nodes as the high-confidence
PIN, and the largest of the five lower-confidence PINs has 25\% more
edges than the high-confidence PIN, i.e., 10,403 of them. This network set has been used in many
existing static NA studies
\citep{GRAAL,HGRAAL,MIGRAAL,MAGNA,LocalVsGlobal,multiMAGNA++}.
When we use the six static PINs as snapshots to form a dynamic
network, we order the six networks from the smallest one in terms of the
number of edges (i.e., one of the highest confidence) to the largest
one in terms of the number of edges (i.e., one of the lowest
confidence). Since each static PIN contains the same
set of nodes, this simulates a dynamic network that is growing as it evolves, with
more and more interactions being added to the network
over time.
When we align the resulting (original) dynamic yeast network to its randomized versions, we find
that just as for the zebra network, DynaMAGNA++'s alignment quality
decreases with increase in the noise level, with respect to both its
objective function (Figure \ref{fig:test2b}(a)) as well as node
correctness (Figure \ref{fig:test2b}(b)), while MAGNA++'s alignment
quality does not change. Further, DynaMAGNA++ again matches more
closely the quality of the perfect alignments than MAGNA++ does with
respect to DynaMAGNA++'s objective function (Figure
\ref{fig:test2b}(a)). Finally, DynaMAGNA++ produces higher node
correctness than MAGNA++ for the lower (meaningful) noise
levels. Thus, dynamic NA is superior to static NA for the yeast
network as well.

\begin{figure}
\centering
\includegraphics[width=0.85\linewidth]{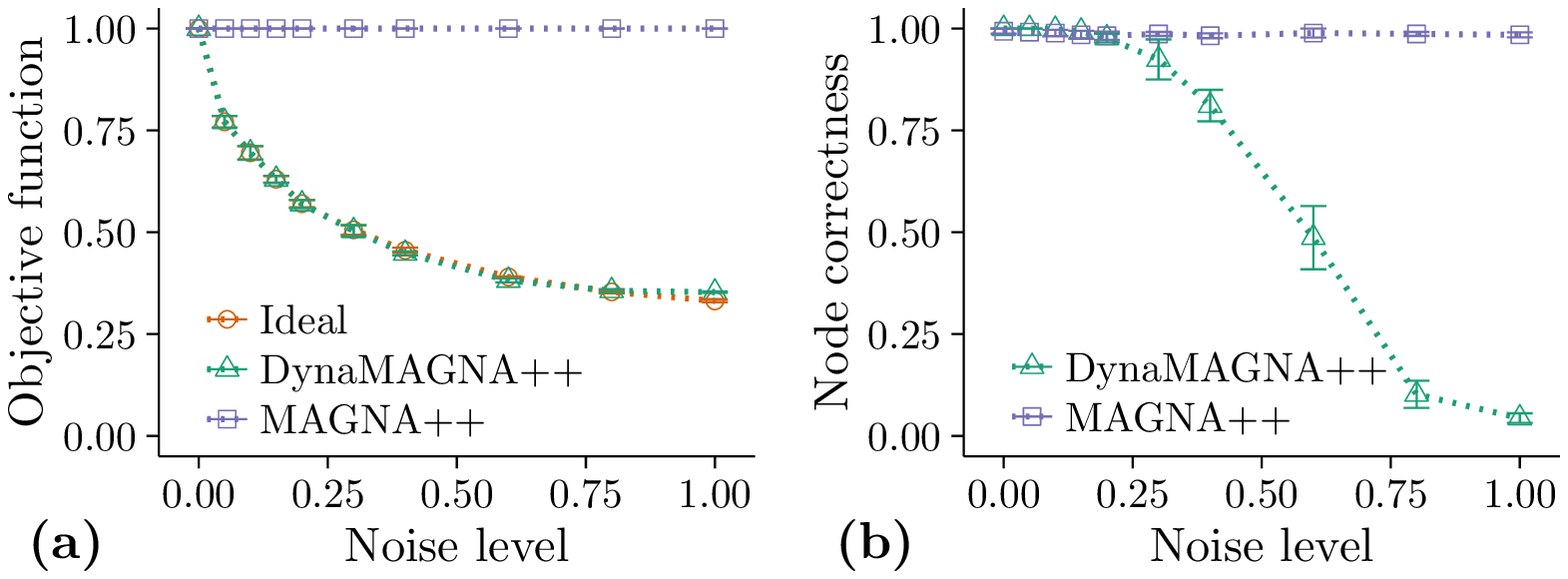}%
    \caption{Alignment quality of DynaMAGNA++ and MAGNA++ for the
      Enron network. The figure can be interpreted in the same way as Figure \ref{fig:test2a}.}
\label{fig:test2c}
\end{figure}

\vspace{0.2em}\noindent {\bf Enron network.}
To demonstrate DynaMAGNA++'s generalizability on
non-biological networks, we continue our evaluation of real-world
networks on
a social network. The original network that we use is the Enron e-mail
communication network \citep{Enron}, which is based on e-mail communications of 184 employees in the
Enron corporation from 2000 to 2002, made public by the Federal Energy
Regulatory Commission during its investigation. The entire two-year time period is
divided into two-month periods so that if there is at least one e-mail
sent between two people within a particular two-month period, then
there exists an event between the two people during that period. There
are 5,539 events in the Enron network. When we align the original
network to its randomized versions, we find that just as for the zebra
and yeast networks, DynaMAGNA++'s alignment quality
decreases with increase in the noise level,
while MAGNA++'s alignment
quality does not change (Figure \ref{fig:test2c}). Further, DynaMAGNA++ again matches more
closely the quality of the perfect alignments (Figure
\ref{fig:test2b}(a)).
So, these results again indicate that dynamic NA is superior to static
NA. Interestingly, for this network, MAGNA++ also
produces high-quality alignments with respect to node correctness for
the low noise levels, just like DynaMAGNA++ does.

\vspace{0.2em}\noindent {\bf Running time.} Recall that the time
complexity of DynaMAGNA++ is linear with respect to the number of
events in the aligned networks (Section \ref{sec:dynamagna}), while the
time complexity of MAGNA++ is linear with respect to the number of
edges in the aligned networks (Section \ref{sec:magna}).  Because
there are typically far more events in a typical dynamic networks than edges in
the flattened version of the dynamic network, and due to the more
involved computations when calculating event conservation, DynaMAGNA++ is expected
to be slower than MAGNA++ (yet, it is this ability of DynaMAGNA++ to
capture detailed temporal event information that makes it superior to
MAGNA++ in terms of accuracy).  A representative runtime for
DynaMAGNA++ and MAGNA++ when the methods are run on eight cores to
align the yeast network to its 0\% randomized version are 1.9 hours
and 0.7 hours, respectively, which makes DynaMAGNA++ 2.7 times slower
than MAGNA++. Yet, this somewhat slower (yet still very practical)
runtime of DynaMAGNA++ is justified
by DynaMAGNA++'s superiority over MAGNA++ in terms of alignment quality.

\vspace{0.2em}\noindent {\bf DynaMAGNA++'s availability.} We implement
a friendly graphical user interface (GUI) for DynaMAGNA++ (Figure
\ref{fig:gui}) for easy use by domain (e.g., biological)
scientists. Also, we provide the source code of DynaMAGNA++ so that
computational scientists may potentially extend the work (available
upon request).

\begin{figure}
\centering
\includegraphics[width=\linewidth]{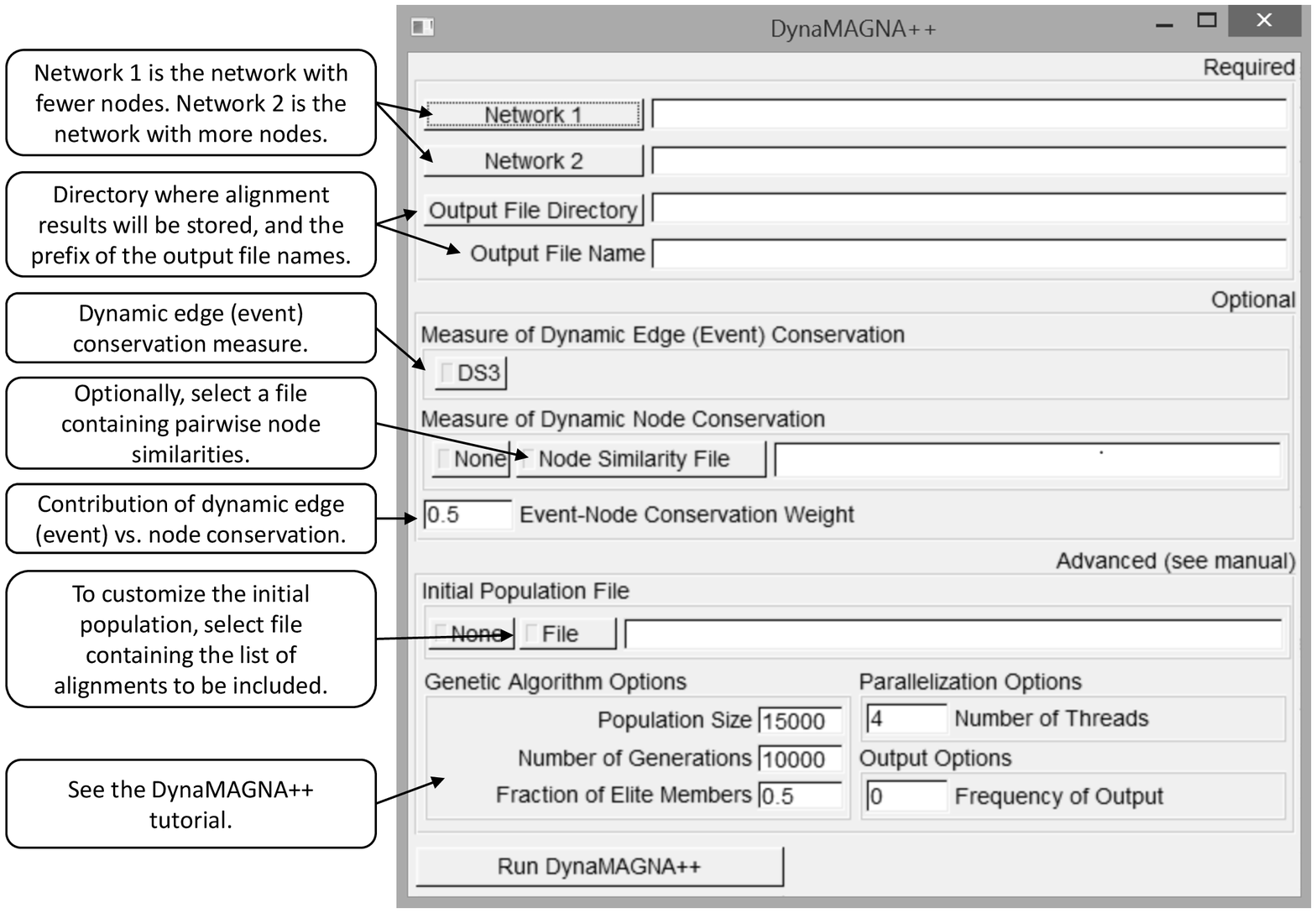}%
\caption{GUI to DynaMAGNA++. The only required parameters are the two
networks to be aligned, and the output directory/file name
information.  While DS$^3$ is the only currently implemented dynamic
edge (event) conservation measure, other future dynamic edge
conservation measures can be easily added.  Any dynamic node
similarity measure can be used by selecting a file containing pairwise
similarities between nodes of the two networks. The $\alpha$ parameter
(Section \ref{sec:dynamagna}) can be set to any desired value.  Other
advanced parameters can also be user-specified.  The default values
are set according to the parameter values used in this work (Section
\ref{sec:dynamagna}).}
\label{fig:gui}
\end{figure}

\section{Conclusion}
\label{sec:conclusion}

We introduce the first ever dynamic NA
method.
We show
that our method, DynaMAGNA++, produces superior alignments compared to its static
NA counterpart due its explicit use of available temporal information
in dynamic network data.
DynaMAGNA++ is a search-based NA method that can optimize any
alignment quality measure. In this work, we
propose an efficient temporal information-based
alignment quality measure, DS$^3$, that DynaMAGNA++
optimizes in order to find good alignments.
DynaMAGNA++ can be extended in two ways: by optimizing future,
potentially more efficient alignment quality measures with the current
search strategy,
or by optimizing its current alignment quality measures with a future,
potentially superior search strategy \citep{FairEval}.

We demonstrate applicability of DynaMAGNA++ and dynamic NA in general
in multiple domains: biological networks (ecological networks and
PINs) and social networks.  Given the impact that static NA has
had in computational biology, as more PIN and other molecular
dynamic network data are becoming available, dynamic NA and thus our
study will continue to gain importance. The same holds for other
domains in which increasing amounts of real-world dynamic network data
are being collected. So, we have just scratched the tip of the iceberg
called dynamic NA.

\paragraph{Funding:}
This work was supported by the National Science Foundation (NSF)
[CCF-1319469] and the Air Force Office of Scientific Research (AFOSR)
[YIP FA9550-16-1-0147].

\bibliographystyle{natbib}
%
%

\bibliography{cone}

\end{document}

%% file: exp_1k_a_objective_function_algorithm_comparison_ss.tex
NA method & AUPR & F-score$_{\mbox{cross}}$ & F-score$_{\mbox{max}}$ & AUROC \\ 
  \midrule
DynaMAGNA++ (E+N) & 0.836 & 0.675 & 0.788 & 0.928 \\ 
  DynaMAGNA++ (E) & 0.551 & 0.400 & 0.750 & 0.878 \\ 
  DynaMAGNA++ (N) & 0.770 & 0.625 & 0.808 & 0.934 \\ 
   \bottomrule

%% file: exp_1k_s_objective_function_algorithm_comparison_ss.tex
NA method & AUPR & F-score$_{\mbox{cross}}$ & F-score$_{\mbox{max}}$ & AUROC \\ 
  \midrule
DynaMAGNA++ & 0.836 & 0.675 & 0.788 & 0.928 \\ 
  MAGNA++ & 0.711 & 0.550 & 0.645 & 0.863 \\ 
   \bottomrule

%% file: main.bbl
\begin{thebibliography}{}

\bibitem[Bayati {\em et~al.}(2013)Bayati, Gerritsen, Gleich, Saberi, and
  Wang]{NApaperBayatiJ}
Bayati, M., Gerritsen, M., Gleich, D., Saberi, A., and Wang, Y. (2013).
\newblock Message-passing algorithms for sparse network alignment.
\newblock {\em ACM Trans. Knowl. Discov. Data\/}, {\bf 7}(1), 3:1--3:31.

\bibitem[Boccaletti {\em et~al.}(2006)Boccaletti, Latora, Moreno, Chavez, and
  Hwang]{ComplexNetworks}
Boccaletti, S., Latora, V., Moreno, Y., Chavez, M., and Hwang, D.-U. (2006).
\newblock Complex networks: structure and dynamics.
\newblock {\em Phys. Rep.}, {\bf 424}(4--5), 175 -- 308.

\bibitem[Crawford {\em et~al.}(2015)Crawford, Sun, and
  Milenkovi\'{c}]{FairEval}
Crawford, J., Sun, Y., and Milenkovi\'{c}, T. (2015).
\newblock Fair evaluation of global network aligners.
\newblock {\em Algorithms for Molecular Biology\/}, {\bf 10}(19).

\bibitem[Duchenne {\em et~al.}(2011)Duchenne, Bach, Kweon, and
  Ponce]{NApaperDuchenne}
Duchenne, O., Bach, F., Kweon, I.-S., and Ponce, J. (2011).
\newblock A tensor-based algorithm for high-order graph matching.
\newblock {\em Pattern Analysis and Machine Intelligence, IEEE Transactions
  on\/}, {\bf 33}(12), 2383--2395.

\bibitem[Elmsallati {\em et~al.}(2016)Elmsallati, Clark, and
  Kalita]{ElmsallatiReview}
Elmsallati, A., Clark, C., and Kalita, J. (2016).
\newblock Global alignment of protein-protein interaction networks: A survey.
\newblock {\em {IEEE/ACM} Trans. on Computational Biology and
  Bioinformormatics\/}, {\bf 13}(4), 689--705.

\bibitem[Emmert-Streib {\em et~al.}(2016)Emmert-Streib, Dehmer, and
  Shi]{FiftyYears}
Emmert-Streib, F., Dehmer, M., and Shi, Y. (2016).
\newblock Fifty years of graph matching, network alignment and network
  comparison.
\newblock {\em Info. Sciences\/}, {\bf 346}(C), 180--197.

\bibitem[Faisal {\em et~al.}(2015a)Faisal, Zhao, and
  Milenkovi\'{c}]{FaisalAging}
Faisal, F., Zhao, H., and Milenkovi\'{c}, T. (2015a).
\newblock Global network alignment in the context of aging.
\newblock {\em IEEE/ACM Transactions on Computational Biology and
  Bioinformatics\/}, {\bf 12}(1), 40--52.

\bibitem[Faisal {\em et~al.}(2015b)Faisal, Meng, Crawford, and
  Milenkovi\'{c}]{ConeReview}
Faisal, F., Meng, L., Crawford, J., and Milenkovi\'{c}, T. (2015b).
\newblock The post-genomic era of biological network alignment.
\newblock {\em EURASIP Journal on Bioinformatics and Systems Biology\/}, {\bf
  2015}(1), 1--19.

\bibitem[Guzzi and Milenkovi\'{c}(2017)Guzzi and Milenkovi\'{c}]{GuzziNA}
Guzzi, P.~H. and Milenkovi\'{c}, T. (2017).
\newblock Survey of local and global biological network alignment: the need to
  reconcile the two sides of the same coin.
\newblock {\em Briefings in Bioinformatics\/}, {\bf doi: 10.1093/bib/bbw132}.

\bibitem[Hayes and Mamano(2016)Hayes and Mamano]{SANA}
Hayes, W. and Mamano, N. (2016).
\newblock {SANA}: Simulated annealing network alignment applied to biological
  networks.
\newblock {\em arXiv\/}, {\bf arXiv:1607.02642 [q-bio.MN]}.

\bibitem[Holme(2015)Holme]{ModernTemporal}
Holme, P. (2015).
\newblock Modern temporal network theory: a colloquium.
\newblock {\em The European Physical Journal B\/}, {\bf 88}(9), 1--30.

\bibitem[Hulovatyy {\em et~al.}(2014)Hulovatyy, Solava, and
  Milenkovi\'{c}]{YuriyLinkPrediction}
Hulovatyy, Y., Solava, R.~W., and Milenkovi\'{c}, T. (2014).
\newblock Revealing missing parts of the interactome via link prediction.
\newblock {\em PLOS ONE\/}, {\bf 9}(3), e90073.

\bibitem[Hulovatyy {\em et~al.}(2015)Hulovatyy, Chen, and
  Milenkovi\'{c}]{DynamicGraphlets}
Hulovatyy, Y., Chen, H., and Milenkovi\'{c}, T. (2015).
\newblock Exploring the structure and function of temporal networks with
  dynamic graphlets.
\newblock {\em Bioinformatics\/}, {\bf 31}(12), 171--180.

\bibitem[Ibragimov {\em et~al.}(2013)Ibragimov, Malek, and Baumbach]{GEDEVO}
Ibragimov, R., Malek, M., and Baumbach, J. (2013).
\newblock {GEDEVO}: An evolutionary graph edit distance algorithm for
  biological network alignment.
\newblock In {\em GCB\/}, pages 68--79.

\bibitem[Kuchaiev and Pr\v{z}ulj(2011)Kuchaiev and Pr\v{z}ulj]{MIGRAAL}
Kuchaiev, O. and Pr\v{z}ulj, N. (2011).
\newblock Integrative network alignment reveals large regions of global network
  similarity in yeast and human.
\newblock {\em Bioinformatics\/}, {\bf 27}(10), 1390--1396.

\bibitem[Kuchaiev {\em et~al.}(2010)Kuchaiev, Milenkovi\'{c},
  Memi\v{s}evi\'{c}, Hayes, and Pr\v{z}ulj]{GRAAL}
Kuchaiev, O., Milenkovi\'{c}, T., Memi\v{s}evi\'{c}, V., Hayes, W., and
  Pr\v{z}ulj, N. (2010).
\newblock {Topological network alignment uncovers biological function and
  phylogeny}.
\newblock {\em Journal of The Royal Society Interface\/}, {\bf 7}(50),
  1341--1354.

\bibitem[Malod-Dognin and Pr\v{z}ulj(2015)Malod-Dognin and Pr\v{z}ulj]{LGRAAL}
Malod-Dognin, N. and Pr\v{z}ulj, N. (2015).
\newblock {L-GRAAL}: Lagrangian graphlet-based network aligner.
\newblock {\em Bioinformatics\/}, {\bf 31}(13), 2182--2189.

\bibitem[Meng {\em et~al.}(2016a)Meng, Crawford, Striegel, and
  Milenkovi\'{c}]{IGLOO}
Meng, L., Crawford, J., Striegel, A., and Milenkovi\'{c}, T. (2016a).
\newblock {IGLOO}: Integrating global and local biological network alignment.
\newblock In {\em Proc. of Workshop on Mining and Learning with Graphs (MLG) at
  the Conference on Knowledge Discovery and Data Mining (KDD)\/}.

\bibitem[Meng {\em et~al.}(2016b)Meng, Striegel, and
  Milenkovi\'{c}]{LocalVsGlobal}
Meng, L., Striegel, A., and Milenkovi\'{c}, T. (2016b).
\newblock Local versus global biological network alignment.
\newblock {\em Bioinformatics\/}, {\bf 32}(20), 3155--3164.

\bibitem[Milenkovi\'{c} and Pr\v{z}ulj(2008)Milenkovi\'{c} and
  Pr\v{z}ulj]{Milenkovic2008}
Milenkovi\'{c}, T. and Pr\v{z}ulj, N. (2008).
\newblock Uncovering biological network function via graphlet degree
  signatures.
\newblock {\em Cancer Informatics\/}, {\bf 6}, 257--273.

\bibitem[Milenkovi\'{c} {\em et~al.}(2010)Milenkovi\'{c}, Ng, Hayes, and
  Pr\v{z}ulj]{HGRAAL}
Milenkovi\'{c}, T., Ng, W., Hayes, W., and Pr\v{z}ulj, N. (2010).
\newblock Optimal network alignment with graphlet degree vectors.
\newblock {\em Cancer Informatics\/}, {\bf 9}, 121--137.

\bibitem[Neyshabur {\em et~al.}(2013)Neyshabur, Khadem, Hashemifar, and
  Shahriar~Arab]{NETAL}
Neyshabur, B., Khadem, A., Hashemifar, S., and Shahriar~Arab, S. (2013).
\newblock {NETAL}: a new graph-based method for global alignment of
  protein-protein interaction networks.
\newblock {\em Bioinformatics\/}, {\bf 29}(13), 1654--1662.

\bibitem[Patro and Kingsford(2012)Patro and Kingsford]{GHOST}
Patro, R. and Kingsford, C. (2012).
\newblock Global network alignment using multiscale spectral signatures.
\newblock {\em Bioinformatics\/}, {\bf 28}(23), 3105--3114.

\bibitem[Priebe {\em et~al.}(2005)Priebe, Conroy, Marchette, and Park]{Enron}
Priebe, C.~E., Conroy, J.~M., Marchette, D.~J., and Park, Y. (2005).
\newblock Scan statistics on {Enron} graphs.
\newblock {\em Comput. Math. Organ. Theory\/}, {\bf 11}(3), 229--247.

\bibitem[Pr\v{z}ulj {\em et~al.}(2010)Pr\v{z}ulj, Kuchaiev, Stevanovi\'{c}, and
  Hayes]{BioNetworkModel}
Pr\v{z}ulj, N., Kuchaiev, O., Stevanovi\'{c}, A., and Hayes, W. (2010).
\newblock Geometric evolutionary dynamics of protein interaction networks.
\newblock In {\em Proc. of the Pacific Symposium Biocomputing\/}, pages 4--8.

\bibitem[Przytycka and Kim(2010)Przytycka and Kim]{NetworkIntegrationDynamics}
Przytycka, T.~M. and Kim, Y.-A. (2010).
\newblock Network integration meets network dynamics.
\newblock {\em BMC Bioinformatics\/}, {\bf 8}(48).

\bibitem[Przytycka {\em et~al.}(2010)Przytycka, Singh, and
  Slonim]{DynamicInteractome}
Przytycka, T.~M., Singh, M., and Slonim, D.~K. (2010).
\newblock Toward the dynamic interactome: it's about time.
\newblock {\em Briefings in Bioinformatics\/}, {\bf 11}(1), 15--29.

\bibitem[Rubenstein {\em et~al.}(2015)Rubenstein, Sundaresan, Fischhoff,
  Tantipathananandh, and Berger-Wolf]{Zebra}
Rubenstein, D.~I., Sundaresan, S.~R., Fischhoff, I.~R., Tantipathananandh, C.,
  and Berger-Wolf, T.~Y. (2015).
\newblock Similar but different: dynamic social network analysis highlights
  fundamental differences between the fission-fusion societies of two equid
  species, the onager and {Grevy's} zebra.
\newblock {\em PLOS ONE\/}, {\bf 10}(10), e0138645.

\bibitem[Saraph and Milenkovi\'{c}(2014)Saraph and Milenkovi\'{c}]{MAGNA}
Saraph, V. and Milenkovi\'{c}, T. (2014).
\newblock {MAGNA}: Maximizing accuracy in global network alignment.
\newblock {\em Bioinformatics\/}, {\bf 30}(20), 2931--2940.

\bibitem[Singh {\em et~al.}(2007)Singh, Xu, and Berger]{IsoRank}
Singh, R., Xu, J., and Berger, B. (2007).
\newblock Pairwise global alignment of protein interaction networks by matching
  neighborhood topology.
\newblock In {\em Research in computational molecular biology\/}, pages 16--31.
  Springer.

\bibitem[Sun {\em et~al.}(2015)Sun, Crawford, Tang, and Milenkovi\'{c}]{WAVE}
Sun, Y., Crawford, J., Tang, J., and Milenkovi\'{c}, T. (2015).
\newblock Simultaneous optimization of both node and edge conservation in
  network alignment via {WAVE}.
\newblock In {\em Proc. of Workshop on Algorithms in Bioinformatics (WABI)\/},
  pages 16--39.

\bibitem[Vijayan and Milenkovi\'{c}(2016)Vijayan and
  Milenkovi\'{c}]{multiMAGNA++}
Vijayan, V. and Milenkovi\'{c}, T. (2016).
\newblock Multiple network alignment via {multiMAGNA++}.
\newblock In {\em Proc. of Workshop on Data Mining in Bioinformatics (BIOKDD)
  at the Conference on Knowledge Discovery and Data Mining (KDD)\/}.

\bibitem[Vijayan {\em et~al.}(2015)Vijayan, Saraph, and
  Milenkovi\'{c}]{MAGNA++}
Vijayan, V., Saraph, V., and Milenkovi\'{c}, T. (2015).
\newblock {MAGNA++: Maximizing Accuracy in Global Network Alignment via both
  node and edge conservation}.
\newblock {\em Bioinformatics\/}, {\bf 31}(14), 2409--2411.

\bibitem[Yavero{\u g}lu {\em et~al.}(2015)Yavero{\u g}lu, Milenkovi{\'c}, and
  Pr{\u z}ulj]{AlignmentFree}
Yavero{\u g}lu, {\"O}., Milenkovi{\'c}, T., and Pr{\u z}ulj (2015).
\newblock Proper evaluation of alignment-free network comparison methods.
\newblock {\em Bioinformatics\/}, {\bf 31}(16), 2697--2704.

\bibitem[Zhang {\em et~al.}(2015)Zhang, Tang, Yang, Pei, and Yu]{COSNET}
Zhang, Y., Tang, J., Yang, Z., Pei, J., and Yu, P.~S. (2015).
\newblock {COSNET}: Connecting heterogeneous social networks with local and
  global consistency.
\newblock In {\em Proc. ACM SIGKDD Int. Conf. on Knowledge Discovery and Data
  Mining\/}, pages 1485--1494.

\end{thebibliography}
